# Reconstructing William Lane Craig's Explanation of Absolute Time Based on Mulla Sadra's Philosophy

Mohammad Sadegh Kavyani[1], Habibollah Razmi[2], and Hamid Parsania[3]

## Abstract

After the advent of the theory of special relativity, the existence of absolute time in nature was rejected within the society of physics. In recent decades, William Lane Craig has endeavoured to offer an interpretation of the empirical evidence that support the theory of relativity while maintaining the concept of absolute time. His interpretation, however, is based upon supernatural presuppositions due to which it cannot be accepted as a scientific argument. Here, after explaining Craig's view, we attempt to reconstruct his explanation for absolute time using the concept of general substantial motion of nature, well-known in Mulla Sadra's philosophy as the most important approach in Islamic philosophy; thereby, proving general time for the natural world. Although Craig considers some evidence from modern physics in his reasoning for absolute time, in this paper, after pointing to some evidence, it is discussed that the approach used here better bridges the gap that exists between the metaphysics and the physics of the argument.

**Keywords** Absolute time, Theory of relativity, William Lane Craig, Islamic philosophy, Mulla Sadra's philosophy, Substantial motion, General time

---

[1] Department of Philosophy of Physics, Baqir al-Olum University, Qom, I. R. Iran; e-mail: kavyani@bou.ac.ir

[2] Department of Physics, University of Qom, Qom, I. R. Iran; e-mail: razmi@qom.ac.ir

[3] Department of Philosophy of Social Sciences, Baqir al-Olum University, Qom, I. R. Iran; Department of Social Sciences, University of Tehran. Tehran, I. R. Iran; e-mail: h.parsania@ut.ac.ir

## Introduction

In classical mechanics, universal absolute time was defined as the measure of the chronology of events and a criterion for determining their priority, posteriority, and simultaneity. This time, also called "duration" [4], was considered to have been uniformly flowing from eternity for all observers and independent of the world's natural motions, like a straight line that extends infinitely on both ends. In modern physics and with the developments after Einstein's theory of special relativity, simultaneity was transformed into a completely relative concept, such that the acceptance of any sort of absolute or general time was rejected (Einstein, 1905, p 897; 1920, pp 25-27).

In the contemporary era, the American analytic philosopher and Christian theologian, William Lane Craig, has offered another interpretation of the empirical evidence supporting the theory of relativity which not only does not contradict the existence of absolute time but is also based on a metaphysical argument that confirms it (Craig, 2008, pp 21-41; 2002, pp 146-147; 2001a, p 247). Here, after introducing and then briefly critiquing Craig's view, an attempt is made to reconstruct Craig's explanation of absolute time based on substantial motion proved in the philosophy of the great Islamic thinker Mulla Sadra.

About Islamic philosophy, it must be noted that while it is commonly thought that the death of Ibn Rushd (Averroës) marked the end of a particular line of Islamic philosophy (usually called the Peripatetic school), there arose many later schools of Islamic philosophy that are still active in the Islamic world (Street, 2008). Among them, the school of Mulla Sadra is one of the most profound and most complete philosophies today and is highly regarded by several Islamic philosophers and thinkers.[5] In the next section, we briefly introduce Mulla Sadra's thought and the theory of substantial motion, which have been used as the founding principles of this study.

---

[4] Newton said: Absolute, true, and mathematical time, of itself, and from its own nature flows equably without regard to anything external, and by another name is called duration. Relative, apparent, and common time, is some sensible and external (whether accurate or unequable) measure of duration by the means of motion, which is commonly used instead of true time; such as an hour, a day, a month, a year (Newton, 1846, p 77).

5 In the Islamic world, after Ghazali and Ibn Rushd (Averroës), philosophical thinking ceased to exist; but with the rise of Mulla Sadra, philosophical thought continued, at least among the Shiites: *"It was due to Mulla Sadra's efforts that philosophy was able to maintain and renew its vigour in Shiite Islam while it was silent in the rest of the Dar al-Islam. Mulla Sadra effected an entire revolution in the metaphysics of being by substituting a metaphysics of existence for the traditional meta physics."* (Corbin, 1993, p 342).

## 1. A Brief Summary of Mulla Sadra's Philosophy

### 1.1. Mulla Sadra's Place in Islamic Philosophy

Mulla Sadra (979 A.H/1571 A.C - 1050 A.H/1640 A.C) is one of the most significant Islamic philosophers after Avicenna. Using new methods derived from the principles of his philosophy— well-known as Transcendent Wisdom (TW) (al-ḥikmat al-muta'āliyah)—Sadra discusses and solves several unsolved problems in Peripatetic philosophy. Sadra's ontology is based on the originality of existence essence aiming at how we can understand the relationship between God, cosmos, and the human state. His theory of cosmology is based on substantial motion (al-harakat al –jawhariyya). Most of the problems of Islamic theology can be solved in a philosophical form through TW. TW comprises all Islamic schools of philosophy, gnosis, and theology, and connects them to each other; yet its approach is different than all of them. Among about several books and treatises, the greatest masterpiece written by Mulla Sadra is "al-Ḥikmat al-Muta 'āliyah fī al-ʾAsfār al-ʿAqlīyat al-ʾArbaʿah" ("Transcendent Philosophy of the Four Journeys of the Intellect") known simply as "Asfar" (Corbin, 1993, p 342; Tabatabai, 1975, pp 97-98; Nasr, 2006, pp 223-24). TW has profoundly affected Islamic philosophy in Persia, Sadra's homeland, and even in the Indian subcontinent (Kalin, 2014, XXV).

In contemporary times, Iranian scholars such as Muhammad Husayn Tabatabai, Morteza Motahhari, and Abdollah Javadi-Amoli, whom we will mention later in this article, are among the successors of Mulla Sadra's intellectual philosophy. The concept of substantial motion (ḥarakat al-jawharīyah) is one of the greatest achievements of Sadra's philosophy and it has a deep influence on the goal of this article: explaining general time for the entire universe.

### 1.2. Mulla Sadra's Theory of Substantial Motion

Aristotle, Avicenna, and other adherents of their school confine motion to only four accidental categories: quality (kayf), quantity (kamm), situation (waḍʿ), and place (ʾayn), and deny motion in the category of substance. Mulla Sadra, however, rejects this idea because the reality of such a moving substance must itself be in motion, because the cause of a phenomenon in motion must itself undergo motion and change, and therefore the substance, which is the cause of a moving object, also has motion.[6] Moreover, time, in Sadra's view, is not an independently existing substance but rather it is abstracted from motion. This makes time one of the dimensions (alongside spatial dimensions) of the real 'temporal' world. Every body then has its own time even if it appears static. Time is abstracted from the change and renewal in the essence of a body and is a form of its actual extension. Thus, matter, which is subject to time, must also be subject to motion. This differs from the Peripatetic view that denotes merely a static conglomeration of spatio-temporal events. In TW, the movement from the potentiality of a thing to its actuality is, in fact, an abstract notion in the mind, while material beings themselves are in continual state of flux perpetually undergoing essential change. Moreover, this substantial change is a property not only of sublunary elemental beings (those composed of earth, water, air, and fire) but of celestial bodies as well (Abdul Haq, 1972; Kalin, 2003).[7] Mulla Sadra also shows through other rational arguments that the natural world, while its multiplicities are real, has a real unity, which indicates the existence of some form of a general time and it also indicates a single creator for it.

In the latter sections, we will derive general time by using substantial motion for the whole of nature.

## 2. Craig's Explanation of Absolute Time

Newton distinguished between metaphysical and physical time. He considered metaphysical (absolute) time to be due to the divine attribute of eternity, which exists independently of physical events. After the advent of the theory of relativity, the belief in absolute metaphysical time was also considered unfounded. Craig considers this approach of physicists to be inadequate. He argues that the Lorentz's interpretation of the empirical evidence for the theory of relativity can be used to revive absolute time, which is rejected in the standard interpretation. For a better explanation of Craig's point of view, let us explain Lorentz's interpretation of the evidence supporting the theory of relativity.

---

6 In TM, the existence of an accident depends on the existence of its substance, and it is impossible to imagine the existence of an accident without the existence of its substance. Therefore, the substance is the cause of the accident.

7 Mulla Sadra has many arguments in this regard that cannot be stated here. One may refer to the mentioned references.

From Craig's point of view, empirical evidence confirming the theory of relativity (such as the Michelson-Morley experiment) can be interpreted in three different ways all of which are empirically equivalent.[8] The three interpretations were proposed by Einstein, Minkowski, and Lorentz are:

A. Einstein's interpretation consists of a three-dimensional formulation of space-time where physical objects endure through time, but space and time are relative to reference frames. In this interpretation, it is useless to believe in a privileged reference frame of time and space. Only the velocity of light (in vacuum) is constant in every reference frame.

B. Minkowski's interpretation: time is practically another dimension of space, and four-dimensional space-time explains the characteristics of physical events[9].

C. In Lorentz's interpretation, on the contrary Minkowski's interpretation, a three-dimensional space-time for spatial objects that endure through a privileged time is accepted. Thus, Lorentz maintains a privileged reference frame regardless of the velocity of a source of light. Here, the relativity of space and time is defined in systems of motion relative to the preferred reference frame (Lorentz, 1934, p 211; Craig, 2008, pp 21-41). The preferred frame in view of Lorentz is the rest frame of the ether which may be considered as the medium for the propagation of the electromagnetic fields.

Craig uses the existence of God and a dynamic theory of time[10] to prefer Lorentz's interpretation over other interpretations. He then argues as follows:

---

8 Craig asserts that a physical theory includes two things: a mathematical formalism and a physical interpretation of that formalism. (Craig, 2002, p 11) If different theories build upon the same formalism but conclude different interpretations this means that they are empirically equivalent.

9 Einstein later accepted this interpretation. (Einstein, 1920, pp 51-53) Although this is a controversial issue. (Brown, 2005)

10 "In the philosophy of time, there are two distinguished types of theories of time; the A-theory (tensed or dynamic theory) of time and the B-theory (tenseless or static theory) of time. Briefly, in A-theory the division of time into past, present and future, and the passage of time is real, whereas in the B-theory this passage is an illusion and past, present and future are merely the relations one would describe an event as temporally located relative to some frame of reference." (McTaggart, 1908, pp 458-459; Dyke 2002, pp 137-145) Craig believes that an A-theory of time is correct (Craig 2001 b, p 81; pp 102-103).

Here, we do not seek to examine Craig's view in detail; even though we accept his critique of the static theory of time, we do not completely agree with a dynamic theory. We accept a middle point that includes both static and dynamic theories of time, the explanation of which is outside the scope of this paper. (Kavyani; Parsania; Razmi, 2022).

1. If God exists and a tensed theory of time is correct, then God is in time.[11]
2. If God is in time, then a privileged reference frame exists.
3. If a privileged reference frame exists, then a Lorentzian interpretation of relativity is correct.
4. God exists and a tensed theory of time is correct.
5. Lorentzian interpretation of relativity is correct. (Craig 2008, p 22; 2001a, p 283)

Craig argues that if God is in time, then the "now" of his metaphysical time demarcates a three-dimensional slice of space-time which is equally "now". This universal frame of reference would thus be privileged so that the events which God knows to be present in it are simultaneous, and absolute time, length, and motion would all be known to God. Rods and clocks in motion relative to it undergo intrinsic contraction and retardation (Craig, 2002, pp 146-147).

Craig's argument for explaining absolute time is based on multiple metaphysical assumptions, which are not acceptable in a scientific argument. These assumptions include: the existence of God, the unity of God, creation and divine knowledge and God's need for a preferred framework in knowledge over temporal beings.[12]

---

11 Another assumption Craig makes in explaining absolute time is the belief in the temporality of God. He describes God to be completely timeless before the creation of the universe, but after it, God is in time. He argues for this using God's knowledge and His causality through time (Craig, 2001 a, p 57; p 114; 1979, pp 65-149). However, as a Christian philosopher, Craig does not consider God's being in time to constitute a change in the Divine Essence. (Craig 2001a, p 271) In this study, we do not seek to critique Craig's argument in detail. But in short, some drawbacks of his explanation are:

1. How can a fixed being not be in time first and then become in time? Indeed, time is nothing but to imagine two subsequent states that cannot be combined. Therefore, the idea that God was timeless and then created time is contradictory (Avicenna, 2009, pp 237-238).

2. The view of Islamic philosophy, which is also confirmed by Islamic traditions (Ibn Babawayh, 2019, p 139), is that God knows the entirety of his creation in full detail before creation, just as he knows them in the same way after creation. Therefore, God knows everything in time, even if limited observers participate in temporality (Mulla Sadra, 1981, v 6, p 190; Kuchnani, Bahrayni, 2006).

3. "Now" is only a hypothetical cut of time that has no external existence because its "length" should be exactly zero, and from nothing, nothing real emerges. Thus, God can't participate in the real world within this time.

12 Stating these problems does not mean that the authors or even Mulla Sadra doubts the existence or the oneness of God; rather, we wish to avoid metamaterial presuppositions in our argument.

## 3. Reconstructing Craig's Explanation for Absolute Time based on the General Substantial Motion of the Universe

Substantial motion of the whole universe is a good tool for explaining general time. In this section, we will present an argument for general time based on the general motion of the universe which is the result of the substantive motion that is well-known in Mull Sadra's philosophy.

### 3.1. Substantial Motion Leads to General Time

Some concepts are relative such that to acquire meaning, they must be compared to other concepts, such as "nearness" and "farness", "aboveness" and "belowness", etc. For example, the place of object *A* may be higher than the place of another object *B* but at the same time be lower than the place of object *C*. In contrast, some concepts are non-relative. The clearest examples of such concepts are essential properties that are abstracted only from the essence of a phenomenon without regard to anything else. For example, human as a concept is intrinsic and non-relative; that is, being a human being is not dependent upon a subject.

Now, with Mulla Sadra's explanation of motion, it becomes clear that every motion due to substantial motion is not relative, rather it is dependent upon the changing essence of a thing, such that a phenomenon is either fixed (its components are together) or variable (its components do not add up). Thus, in establishing permanence and change, there is no need for a comparison to another. In other words, the objective assertion regarding a thing does not change with varying subjects (Javadi Amoli, 2015, v 12, pp 356-357). Therefore, time, as the "measure" of motion, is also non-relative and objective, and its quantity has nothing to do with its subject (Mulla Sadra, 1981, v 3, p 414; Fanaei Eshkevari, 2007, pp 40-41).

By substantial motion, Mulla Sadra means that the reality of a material substance is nothing but its motion. It is not the case that matter is an essence upon which motion becomes an accident, but rather motion is the exact essence of matter. Therefore, in the material world, one cannot imagine a motionless substance. This general motion does not depend upon the subject and pertains to all essences within the material world (Mulla Sadra, 1981, v 7, p 298; Fanaei Eshkevari, 2007, p 37)[13].

---

13 Islamic narrations also affirm this view of Mulla Sadra (Ibn Babawayh, 2019, p 261).

As is seen in Subsection 1.2, from Mulla Sadra's viewpoint, although motion and time are two different concepts in our mind, they aren't different in the real world. Time is the "measure" of motion; so, if motion is common to the entire material world, its time is also common; that is, a common time will apply to the entire natural world. This time, according to the previous discussions, doesn't depend on its subject; it is absolute and non-relative. It can also be said that temporal priority and posteriority between two objects is possible only if a common time is realized between them, and since the whole natural world is in motion—due to substantial motion—a common time must also flow through the whole universe (Tabatabai 2007, v 3, p 843; Motahhari, 2014, v 3, pp 219-220).

Morever, the general motion and the general time introduced here lead to a form of unity between the components of nature; because, any general time and motion must also have a common locus (in which motion and time occur), otherwise it is not possible to abstract from two entirely distinct essences a common reality which describes their nature (Tabatabai, 2007, v 3, p 844; Motahhari 2014, v 3, p 221; Fanaei Eshkevari, 2007, p 38).

### 3.2. Do the Experimental Evidence for the Theory of Relativity Reason on Absence of an Absolute Time?

As we know, based on the theory of relativity, there is no preferred frame amongst inertial frames, and no absolute time or dimension can be attributed to objects; rather every onlooker reports these amounts to be different from another based on his/her velocity (Einstein 1920, pp 22-29). This is against the viewpoint on time in classical mechanics and traditional philosophy, wherein spatial and temporal dimensions of an event are invariant and independent of any observer. How can we resolve this difficulty? In response, it should be noted that an inherent and absolute phenomenon may appear differently to different observers. This view is quite acceptable from the perspective of Islamic philosophy because different observers do not have to see an object equally.[14] However, the interpretation of some physicists regarding the theory of relativity shows that they are skeptical about objects having precise dimensions in and of themselves, such that even a stationary frame relative to an object (a proper frame) does not indicate its correct dimensions.

14 Such phenomena have been observed even before the theory of relativity in classical mechanics such as the Doppler effect.

*"A rod in Einstein's theory has various lengths according to the point of view of different observers. One of these lengths, the statistical length, is the greatest, but this does not make it more real than the others… Exactly corresponding remarks apply to the relativity of time... the proper time seems longer. Here, too, it is meaningless to ask "What is the real duration of an event?"* (Born, 1927, pp 213-214)

This viewpoint is also posited about temporal priority and posteriority relative to one another, such that between two distant temporal events, one cannot say which truly precedes the other.[15] Now, it should be known that the relativistic idea of the dependency of the absolute cognition of time upon the observer's status is similar to Immanuel Kant's view of the category of time[16] (Kant, 1919, p 88), and some great scientific scholars, like Kurt Gödel, have also confessed to this point (Gödel, 1995, p 247). However, this approach to the analysis of the reality of time in Einstein's theory of relativity—similar to Kant's view where the knowledge of the reality of things is dependent upon the subject—denies the possibility of identifying external objective truths that exist independent from any observer. The logical result is the denial of existence by itself *('Fi Nafseh'*) of objective truth and a complete rejection of realism in favor of idealism. A better explanation regarding the evidence for the theory of relativity is presented in the next section.

Although physicists generally believe that experiments like Michelson-Morley's have denied the existence of an absolute frame of reference (ether) (Resnick, 1968, pp 18-34), however, as Lorentz believed, all of the results of such experiments can be explained without discarding the absolute frame[17] (Lorentz, 1934, p 211; Craig, 2008, pp 13-16). The theory of Lorentz ether, despite its equivalence to the standard theory of relativity at an empirical level, has not been accepted by the physics society; this is because it is not as "elegant" and 'simple" as Einstein's theory—Lorentz interpretation is much more complex (Rindler, 2006, pp 10-12). However, 'simplicity" and "elegance" by themselves are not sound criteria for accepting or rejecting a theory[18]. Moreover, and fundamentally more important, discarding the theory of the ether has been based on the notion that non-observable means non-existent; whereas not finding something is not a proof for of its non-existence.

---

15 Even if the proper time (the time measured in the frame of reference related to the object within it) of an event could be introduced as the absolute and invariable time for assessing the real quantity of its progress (unlike Max Born's viewpoint mentioned above), no general and absolute time can be defined in the standard theory of relativity for various events that are in different proper frames, because there is essentially no way for introducing absolute simultaneity between distant frameworks.

16 Of course, from Einstein's viewpoint, the observer's physical position influences knowing the time of an event, but Kant takes the observer's senses as being effective therein.

17 Although Galilean transformations cannot explain the results of experiments such as Michelson-Morley's experiment by upholding the existence of an absolute frame, all aforesaid experiments can be justified through Lorentz's transformations. Acceptance of these transformations does not entail the denial of the absolute frame because the movement of an observer relative to an absolute frame causes discrepancies in the measuring apparatuses for time and space. This influences the calculations of the spatial and temporal dimensions of a physical event (Bell, 1986, pp 49-50; Kuhne 2002).

All the way, although Craig himself paid enough attention to the above-mentioned points (Craig, 2002, pp 132-37), his argument needs more strong physical (not purely metaphysical) evidence.

### 3.3. The Properties of General Time and the Origin of its Abstraction

When asking about the time of an event (category when[19] in Aristotelian categories), the answer should be provided with general time, not the special time of the event. For instance, when asking the question "When did event *A* happen?", times other than the special time of event *A* occurrence should be provided in an appropriate answer, such as "last year"[20] or "yesterday". This time should be based on the oldest and most consistent motion in nature and it has to be completely non-relative and constantly in movement so that it can be the proper basis for assessing the motion of everything in the world of nature in terms of precedence, subsequence, decrease, and increase between them (Mulla Sadra, 1981, v 4, p 220; Javadi Amoli, 2015, v 14, pp 28-29). As for the measure of general time commonly called "year", "hour" or 'second", there are various philosophical perspectives:

---

18 As some scholars of the philosophy of science have argued, the justification for 'simplicity" does not follow a single global criterion; but it is local. That is, it can be legitimate in one matter, and illegitimate in another (Sober, 1994, pp 136-157).

19 There is a difference in natural philosophy between the concept of time or occasion (time as an extended non-static quantity) and the concept of "when" and "what time" (category when).

20 Without a comparison to another movement, "year" only shows the quantity of the Earth's movement around the Sun, not the quantity of phenomenon A's movement.

A. Peripatetic philosophers know this time as the quantity of celestial positional motion[21] (Avicenna, 1996, p 105).

B. Some Muslim thinkers consider time, similar to Newton, as a continuous immaterial substance that flows uniformly outside the natural world (Razi, Abo Hatam, 2002, pp 14-15). Some philosophers have termed this time dahr, and similar usage of this term can be found in the words of some Muslim theologians (Sheykh Mofid, 1993, pp 66-67; Razi, Fakhr al-Din, 1987, p 91).

C. From the perspective of Mulla Sadra, this time is neither the quantity of the positional motion of celestial spheres nor is it unrelated to motion within the universe, rather it is understandable based on substantial motion of the celestial spheres.[22] In continuation and completion of this perspective of Sadra, contemporary Neo-Sadraeans consider time as flowing through the entirety of the natural world due to the common substantial motion of everything in nature. This common motion itself serves as proof for a type of real unity between the components of nature, because time is the measure of motion and this motion signifies a sort of real unification between the components of nature[23] (Tabatabai, 2007, v 4, pp 1249-1250; Motahhari, 2014, v 3, pp 218-220; v 4, pp 231-232; Javadi Amoli, 2015, v 15, p 285). More clearly, M. H Tabatabai, as the head of these Neo-Sadraeians, says:

*"When there is a temporal precedence relative to a movement or a mover, this requires the actualization of a common time between them, which necessitates a common movement for all of them, and thus a common matter shared by them"*. (Tabatabai, 2007, v 3, p 844)

---

21 To Peripatetic philosophers, special time did not matter so much for they did not believe in substantial motion) and (general) time was also only obtained based on the celestial positional motion of the spheres.

22 From the perspective of Mulla Sadra, other natural phenomena do not have the required persistence and arrangement for determining general time (Mulla Sadra, 1981 with Sabzevari's commentary, v 3, p 116, footnote 1).

23 Because common substantial motion requires a common substantial subject upon which motion occurs; however, entirely distinct things cannot have a real relation in this regard.

Now, it has to be noted that perspective A above is based on Ptolemaic cosmology which is not devoid of error according to modern cosmology and the ideas mentioned regarding the conditions of general time (Mulla Sadra 1981 with Tabatabai’s commentary, v 3, p 117, footnote 2). Furthermore, the common aspect of general time in this perspective is not based on any reality but is merely mental; since, from the perspective of the general public, celestial motion is more general with respect to the other movements and, therefore, this perspective cannot be the premise for a real philosophical theory (Javadi Amoli, 2015, v 14, p 49).

General time in perspective B is independent of the motion of the world of nature; this is not in accordance to the philosophical definition and terminology of time. Thus, this time does not have an external origin of abstraction (Ibid, pp 282-284).

Therefore, C is the most appropriate of these perspectives because time in this perspective has a real source of external abstraction and its common aspect is not merely mental but a real ontological one. What is understood from Mulla Sadra on the origin and reality of time (Mulla Sadra 1981, v 3, pp 138-14) is that time’s origin of abstraction is not a separate motion apart from the motion of natural objects (including their special motions which themselves are the origin of the abstraction of special time). Indeed, the global motion of all parts of the universe is due to the collective and co-moving motion of all its constituents:

*“The entire natural world enjoys a form of real unity even with the multiplicity it possesses. With all its components, the world is similar to a moving convoy and has a single movement, and this is exactly from where common time comes about.”* (Motahhari, 2014, v 3, p 221)

This issue even has implications in the affirmation of Divine Unity, because the real unity of motion of the components of the universe shows a necessary concomitance among them. If there is a necessary concomitance among all components of the world, they should be effects of a single cause that has created the necessary concomitance among them. Islamic narrations also affirm this view of Mulla Sadra because the unity and connected plan (ʾitteṣāl al-tadbīr: Ibn Babawayh, 2019, p 250) in the natural world indicates the unity of its creator (Mulla Sadra, 2004, p 37).

In summary and diagrammatically:

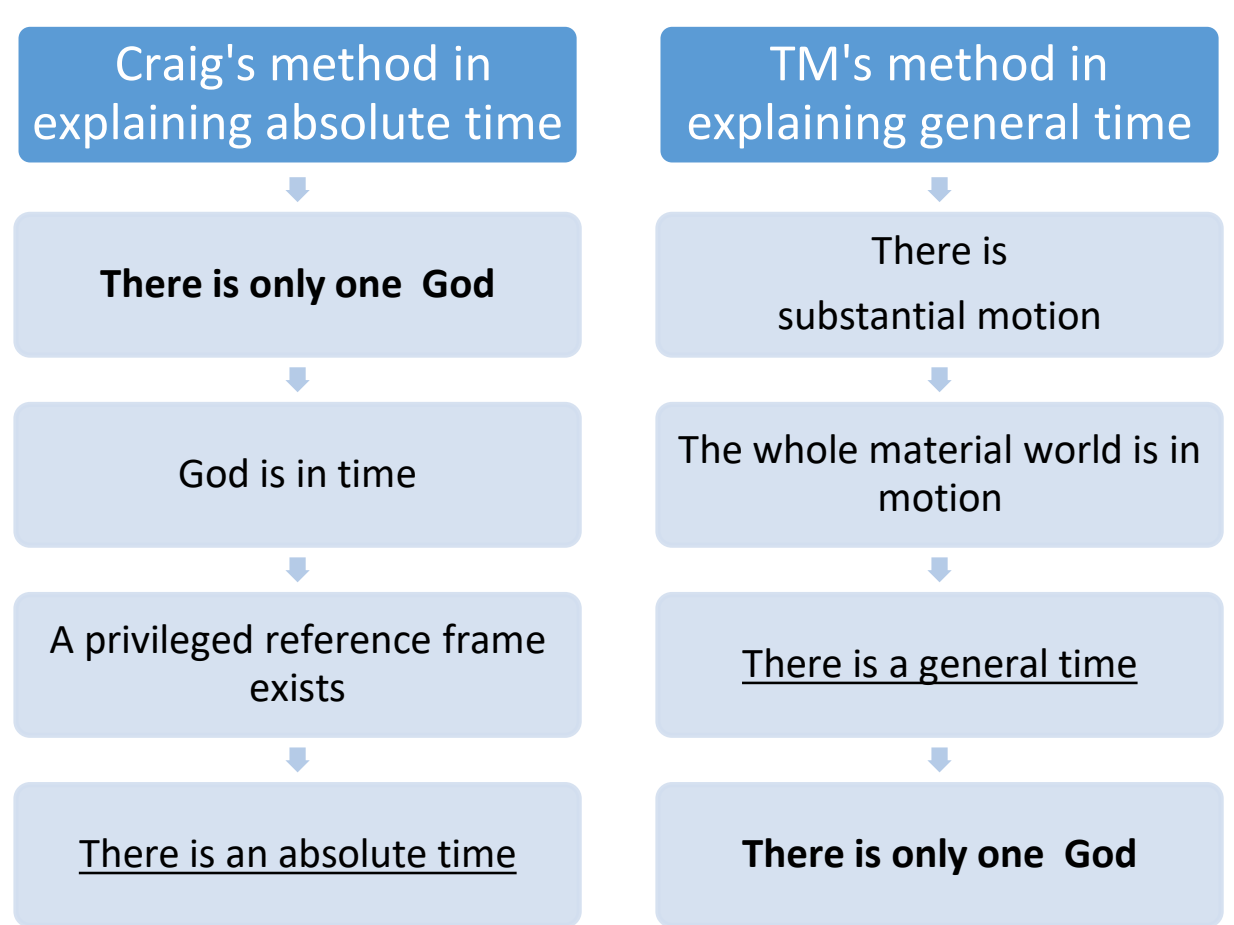

*Diagram 1: Comparison between TM and Craig's Methods*

The difference between TM's and Craig's methods is less and more the same as the difference between Aristotle's and Plato's methods about the world, which has been reflected in Raphael's painting (in Aristotelian thought, just in opposite to Platonic thought, understanding the supernatural world begins with the knowledge of natural world)[24].

## 4. Evidence from Modern Physics Indicating the Realization of the Non-relative and General Time in the Universe

In this section, despite Craig's approach which is based on some supernatural remarks as the existence of God, we want to use a more natural argument using some evidence from modern physics in realizing the so-called general time. In more detail, although Craig also uses some evidence from modern physics in justifying his result on absolute time, we try to use some witnesses from modern physics for both the premises and final results of TW's explanation of general time to be more believable for physicists.

### I. The Whole Material World is in Motion

According to the well-known Heisenberg uncertainty principle in quantum theory and its advanced relativistic form known as quantum fields theory, all objects in the universe are

24 It should be mentioned that although TW's approach used here to general time is from the natural world to the supernatural world, this isn't always true for other arguments in TW.

in an intrinsic (non-relative) motion due to a continual fluctuation in the vacuum field (quantum vacuum). This fact that all the components of the universe are constantly moving confirms the general motion of the universe considered here (Dirac, 1927, p 243; Fock, 1932, pp 622-647).

II. **There is a Unity Between All the Components of Nature**

Despite what introduced by the theory of relativity in separating the universe as local "islands" (the locality principle), today, based on experimentally confirmed results of quantum theory through Bell's theorem, it is known that nature has a nonlocal character technically named as the entanglement property between the different components of the world. As d'Espagnat remarks in this regard: "We *may safely say that non-separability is now one of the most certain general concepts in physics*." ( Kafatos; Nadeau, 1981) This necessitates a kind of real unity among all parts of the universe. This unity can lead to things in common, like the common/general time.

III. **General Movement Introduces General Time**

As is seen in Subsection (3.1), one of the most important premises of the explanation under consideration (TM) corresponds to the whole general motion of the universe. Here, we want to use the well-known cosmic time from modern cosmology as evidence in this regard; we will pay more attention to it in the next subsection.

## 4.1. Cosmic and/or Absolute/General Time

Astronomers have already estimated that the universe is 13.8 billion years old. As we know, the universe's age is calculated using the cosmic time, which is a general quantity corresponding to the overall motion of the universe. The cosmic time can be considered as evidence from modern cosmology for the realization of absolute time. The cosmic time does not express the amount of the peculiar motion of a part of the universe rather it signifies the co-moving of all the components of the universe from the beginning of its formation till now. This is seen in a simple relation known as Hubble's law, which is the result of several observational data and it leads to the overall expansion of the universe which, itself indicates the common motion of the various parts of the universe[25]. This does not mean that the universe components are distancing away from one another towards outside the universe. Rather, it means that the entire components of the universe are moving with respect to each other based on a universal relation[26]. The cosmic time is non-relative and it does not belong to a specific body. This issue can also be explained by the theory of general relativity[27] which is known as the theoretical basis for modern cosmology. More technically, the Friedmann equations[28] governing the time evolution of the universe are written in terms of general cosmic time. These equations enable us to understand how the universe expands and to find its age in absolute and non-relative terms (Rindler, 2006, pp 397-406).

### 4.2. Some Points on the Absoluty of the Cosmic Time

Recently, introducing cosmic time as absolute time has been criticized (Read; Qureshi-Hours, 2021; Callendar; McCoy, 2021); the critics don't accept the cosmic time as an actual universal time especially because they use relativistic (both special and general gravitational) considerations about which it will be explained more as in the following:

A. Hubble used redshifts of distant galaxies to reach his universal expansion law; it is clear that there is a "philosophical" assumption behind the apparently simple Hubble law. The cosmological principle based on which cosmologists generalize local results to all the universe is the basis for this assumption. Indeed, what has been observationally confirmed is that the universe is isotropic and homogeneous at large scales (the cosmological principle) for distance scales up to hundreds of billions of light years around us and what makes cosmologists extend this result to all the universe is based on symmetry, simplicity, and beauty which are all philosophical than physical assumptions. The basic observational support for the overall expansion of the universe and the introduction of the cosmic time is the cosmological redshift which, is different from the local relativistic redshifts. As pointed out in 4.1, cosmic time is a global/universal time corresponding to all the universe as a (one) entity and thus the local gravitational redshift should not be confused with such a universal time.

---

25 Hubble's law is expressed in the following form: $V=H.D$ wherein $V$ is the speed in which cosmos components get away from each other; H is the Hubble's constant and $D$ is the distance between the components of cosmos. Although this law has currently undergone nonlinear revisions, it is still realized as the global law governing the entire components of cosmos.

26 According to what was mentioned in the previous section, if there is no true unity existent between cosmos components, it cannot become the origin of such a unique issue's abstraction.

27 Based on this theory, the space-time curvature formed by the presence of matter is in contradiction to the equivalence of different observers (Gödel, 1990, pp 203-204) and in support of the existence of a sort of Ether (Einstein, 1922, p 23).

28 The Friedmann equations have been derived from Einstein's field equation considering the homogeneity and isotropy of the universe on a large scale by means of Lemaître–Robertson–Walker metric.

B. There is also good evidence in cosmology for the realization of the absolute framework. Cosmic background radiation (CMBR) is a piece of this evidence. What was not found by the Michelson-Morley experiment (the speed of the Earth relative to the ether) was later found by observing CMBR by several satellites from COBE to WMAP. Based on the CMBR data, it can be seen that the local cluster of our galaxy (the cluster of the galaxy of which the Milky Way is a part) is moving at a speed of $627 \pm 22$ km / s relative to the background radiation framework (Kogut and et al, 1993). This can be considered as a real case for this fact that when relative phenomena are compared there should be a real absolute phenomenon for this comparison which itself (here CMBR) is really observable. Many apparent "phenomena" and paradoxes like time dilation, Putnam's argument for the negation of presentism (Putnam, 1976), the Andromeda paradox, and the grandfather paradox are somethings as the above-mentioned explanation on CMBR which means that, although one deals with relative times in them, this cannot be considered against a true real background time. In other words, the role of the general cosmic time with respect to already known relative times in theory of relativity is less and more the same as the role of CMBR and the different local parts of the universe. It should be noticed that the cosmic time is a "measure" of the overall expansion of the universe which is an observable really confirmed objective phenomenon completely irrespective of the local peculiar maybe even relative and/or contractive motions of different parts of the universe. Very shortly and in summary, if one considers the cosmic time as an illusive subjective time, then, he/she will arrive at this result that the universe expansion in modern cosmology is a mental nonfactual phenomenon!

C. It should be mentioned that the general time considered here isn't completely the same as Newtonian absolute time. In Newtonian view, time is separated from matter; while general time "depends" on the matter (motion) in the universe. So, the challenges (as those stated by Gödel (Gödel, 1990, pp 202-207) and Earman[29]) on the Newtonian absolute time cannot be necessarily applied to general time[30].

### 4.3. Which explanation is more believable to physicists?

Here, we want to explain a little more about the advantage of the TM view relative to Craig's one about new achievements of modern physics. To do this, based on Islamic philosophical (TM) discussions stated in section 3, let consider specific physical characteristics for a general/absolute time and then compare it with the evidence of modern physics (which has been discussed before). The main character of TM's explanation is based on an overall non-relative co-moving motion for all nature which necessitates a sort of actual unity of the whole components of it. The cosmic time has a key role in this comparison; because it better represents the required characteristics. Although Craig has used the cosmic time as evidence from modern cosmology, his statement is based on the essence of God (Craig, 2001 a, pp 202-210) without stating clear characteristics and physically testable requirements for an absolute time. But, TM's explanation is based on three main characteristics: the intrinsic (non-relative) motion of all nature; unity of all the parts of the world; and, the common motion of all parts of the universe. As was seen in Subsections 4.1 and 4.2, There is evidence for all these characteristics in modern physics. One can utilize continual fluctuations in the quantum vacuum field in support of the intrinsic (non-relative) motion of all objects and the non-local (global) character of entangled particles in the quantum world in support of the unity of nature; and most importantly, the cosmic time related to the co-moving motion of all the components of the universe. These proximities[31] are enough to establish better compatibility of the evidence of modern physics with the view of TW relative to Craig's explanation. Moreover, the Newton's absolute time defended by Craig cannot be supported by physically testable evidence since it is a perfect metaphysically unrealized concept; this is because the Newton's time is an absolute "quantity" regardless of the physical matter.

---

29 Some researchers have stated that the so-called "hole argument" corresponding to general theory of relativity is in conflict with the ontological requirements of presentism (Balashov; Janssen, 2003, p 342; Earman, 1986, Ch. 9); but of course, from other viewpoints:

1. The dynamical form of general relativity is often called "geometrodynamics" in which, like Lorentz's interpretation, space-time has a "3+1" dimensional formulation and it isn't in conflict with presentism (Barbour, 1999, p 167);
2. Although Einstein expected his theory of general relativity had been 'strongly" Machian, the ultimate field equations didn't satisfy his expectation (it is well-known that the solution of GR field equations for zero value of energy-momentum tensor is non-trivial). It seems the root of Einstein's expectation referred to this point that he wanted to solve the "hole argument" by accepting Mach's principle.
3. It should be mentioned that even if the "hole argument" is correct and thus Craig's explanation of absolute time is failed, general time presented in this paper is not a substance independent of natural movements.

30 In any case what makes absolute time and general time related in the reconstruction of Craig's explanation in this paper is their non-relativity and their criterion for refuting the claim of the theory of relativity in the relativity of events.

In summary and diagrammatically:

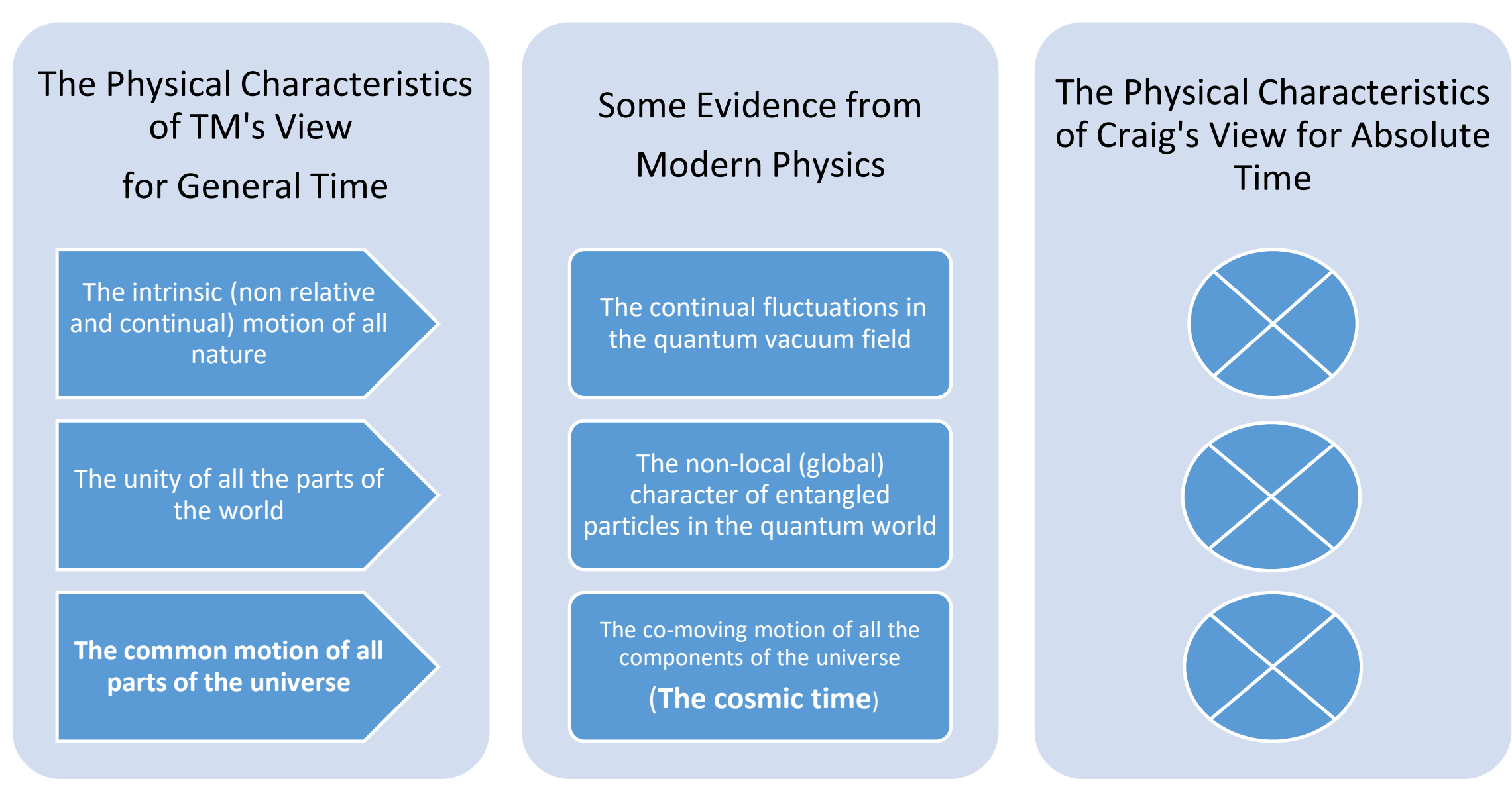

*Diagram 2: Comparison between TM and Craig's views about some evidence from modern physics*

## Conclusion

Mulla Sadra and William Craig are two great philosophers of the East and the West who have important views on the subject of time. In this article, we have reconstructed Craig's views on absolute time with the philosophical foundations of Mulla Sadra. Using his own philosophical and theological principles, such as the temporal existence of God, William Craig revived the absolute time of Newtonian physics, which was rejected by Einstein's theory of relativity. Instead of Einstein and Minkowski's interpretations, he uses Lorentz's theory of relativity to explain this claim. Nevertheless, his particular theological explanation may not be acceptable to physicists because it is based on metaphysical and supernatural assumptions. In this paper, using Substantial motion of the whole of nature known in the philosophy of Mulla Sadra, a general absolute time that can be explained based on testable considerations is introduced. The main features of our explanation are based on a general non-relative motion for the entire natural world, which necessitates a kind of real unity for all the components of the universe.

[31] It is clear that one should not expect a complete correspondence.

As discussed in Section 4, one can utilize some well-known evidence from modern physics in support of an overall motion and unity of the natural world as the intrinsic (non-relative) motion of all objects due to a continual fluctuation in the vacuum field (quantum vacuum), the non-local (global) character of entangled particles in the quantum world, and the cosmic time related to the co-moving motion of all the components of the universe which can be considered as the most important proof. Among these, both because of the great importance of the cosmic time and because this evidence has been more under consideration by the critics who don't accept it as an absolute time, in Sections 3 and 4, it has been explained about this fact that although this time is in support of general time introduced in this paper, it isn't as the same as Newton's absolute time or even what introduced by Craig.

In summary, we have used some information from modern physics to test the coherence of Craig's explanation. We reveal that Craig's work is based on some supernatural pre-assumptions that do not align with modern physics. Furthermore, our explanation is aligned with modern physics while also retaining Craig's interest to explain the unity of the universe's Creator.

*The authors claim there is no any conflict of interest.*